\documentstyle[11pt,paspconf,epsf]{article}

\begin{document}

\title{Evolutionary Models of White Dwarfs with Helium Cores}

\author{T. Driebe\altaffilmark{1} and D. Sch\"onberner}
\affil{Astrophysikalisches Institut Potsdam, An der Sternwarte 16, 
       D-14482 Potsdam, Germany}
\altaffiltext{1}{Present address:  Max-Planck-Institut f\"ur Radioastronomie, 
                                   Auf dem H\"ugel 69, D-53321 Bonn, Germany}      
       
\author{T. Bl\"ocker}
\affil{Max-Planck-Institut f\"ur Radioastronomie, Auf dem H\"ugel 69, 
       D-53321 Bonn, Germany}
        
\author{F. Herwig\altaffilmark{2}}
\affil{Astrophysikalisches Institut Potsdam, An der Sternwarte 16, 
       D-14482 Potsdam, Germany}
\altaffiltext{2}{Present address: Professur Astrophysik, Universit\"at Potsdam, 
                                  Am Neuen Palais 10, D-14469 Potsdam, Germany}

\begin{abstract}
  We present seven evolutionary tracks for low-mass white dwarfs  with 
  helium cores, ranging in mass from 0.179 to 0.414~M$_{\odot}$.  We generated
  the pre-white dwarf models
  from a 1~M$_{\odot}$ sequence extending up to the tip of its
  red-giant branch by applying high mass-loss rates at appropriate positions, 
  and we followed their  evolution across the
  Hertzsprung-Russell diagram and down the cooling path. 
  
  We discuss the internal structures and cooling properties of these new models
  and compare them  with those of recently published models for 
  low-mass white dwarfs which are based on simplified initial 
  configurations. We also demonstrate that our new models seem to remove the 
  apparent discrepancies between the characteristic ages of millisecond pulsars
  and the cooling ages of their white dwarf companions.
\end{abstract}



\section{Introduction}

  The interest in
  low-mass white dwarfs has increased recently because they appear frequently 
  as a binary component, especially in several millisecond pulsar systems. 
  Detailed  evolutionary calculations of possible binary scenarios existed only 
  for  isolated cases or  limited mass ranges with
  $ M > 0.2$~M$_{\odot}$ (Kippenhahn et al.\ 1967, 1968, Refsdal \& Weigert 1969,
  Giannone et al.\ 1970, Iben \& Tutokov 1986, Castellani et al.\ 1994). Only
  recently also calculations for  $ M < 0.2$~M$_{\odot}$ have been presented
  (Alberts et al.\ 1996, Sarna et al.\ 1998). 
  
  The need for more extended sets of white dwarf models with helium cores prompted
  several authors to generate them from ad hoc assumed, simplified
  starting configurations which appear not to be consistent with 
  evolutionary considerations (Althaus \& Benvenuto 1997, Benvenuto \& Althaus 
  1998, Hansen \& Phinney 1998). These calculations are based on the implicit
  assumption that the contraction time to a {\em real\/} white dwarf structure
  is short compared to the cooling time itself. Since also the size of 
  any unprocessed hydrogen-rich envelope can only be guessed, no definitive
  statement on the importance of residual hydrogen burning can be made. 
  
  Thus, we felt it necessary to provide a grid of evolutionary models for 
  low-mass white dwarfs with structures being as consistent as possible with their expected
  evolutionary history, which can be used with
  confidence for interpreting observational data. In our study we aimed at addressing 
  the following questions in a systematic way:
\begin{itemize}   
    \item   How large are the masses of the outer, still unburned hydrogen-rich 
            envelopes on top of the helium cores? Since the size of a white 
            dwarf depends critically on the mass content of its unprocessed 
            envelope, this question is closely related to the mass-radius 
            relation of white dwarfs.
    \item   Can the cooling properties of low-mass white dwarfs be reconciled 
            with estimated spin-down ages of millisecond pulsars?
    \item   Are simplified model calculations useful in interpreting 
            observational data?
\end{itemize}

\section{The evolutionary computations}

   We used an evolutionary code  with  the following 
 basic input physics (Bl\"ocker 1995):
          Nuclear burning was accounted for by a nucleosynthesis network 
          inclu\-ding 31 isotopes and 74 reactions up to carbon burning.
          Radiative opacities were taken from Iglesias et al.\ (1992), 
          complemented with those of Iglesias \& Rogers (1996) and 
          Alexander \& Ferguson (1994) for the low temperature range, 
          all for  $ (Y,Z) = (0.28, 0.02)$.
          Convection was treated according to the mixing length theory. 
          The mixing length parameter was chosen to  $\alpha  = 1.7$, 
          calibrated by computing a solar model.
          Coulomb corrections of the equation of state have been taken from 
          Slattery et al.\ (1982).

  The outcome of the mass transfer in a close binary system was simulated in the
  following simple way:  the evolution of a 1 M$_{\odot}$ model 
  was calculated up to the tip of the red  giant branch (RGB),
  and depending on the desired final mass, high mass loss
  was switched on at the appropriate positions. When the model started to leave 
  the RGB, mass loss was virtually switched off (cf.\ Iben \& Tutukov 1986, 
  Castellani et al.\ 1994). More details of our calculations are given in 
  Driebe et al.\ (1998).

  Our method ensures that these red-giant remnants (= pre-white dwarf models)
  have a structure which is
  consistent with their previous  evolutionary history, with an electron 
  degenerate helium core  and a (mainly) unprocessed envelope.
  The following evolution of the models across the Hertzsprung-Russell diagram 
  and down the cooling path depends only on their actual structure (and 
  mass-loss for larger luminosities), and not on  the details 
  of previous heavy mass-loss episodes during the supposed binary evolution,
  provided the mass donor regains its thermal equilibrium before it finally 
  shrinks below its Roche lobe.
\begin{figure}[th]                      
\plotone{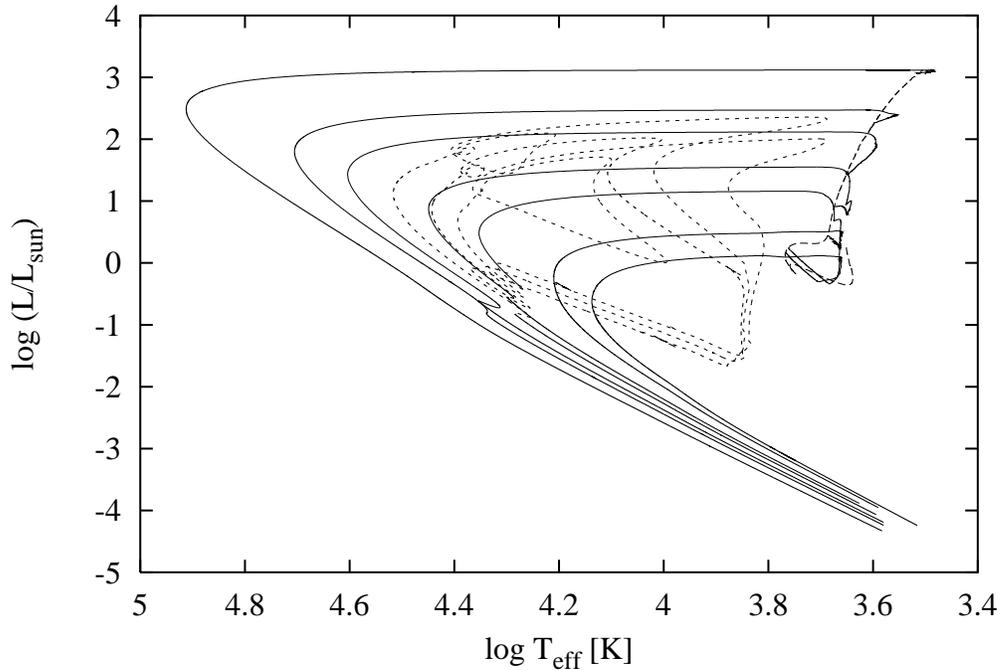} 
\caption{ \label{fig1}
Hetzsprung-Russell diagram with complete evolutionary tracks of RGB remnants 
with different masses (from top: 0.414, 0.331, 0.300, 0.259, 0.234, 
0.195, 0.179~M$_{\odot}$). The long-dashed curve shows the
evolutionary track of the 1 M$_{\odot}$ model we used for creating
 the remnants by mass loss. 
The short-dashed loops outline the very rapid redward excursions of the 0.259 and 
0.234~M$_{\odot}$ models caused by hydrogen shell flashes.  
        }
\end{figure}
  
  Fig.~\ref{fig1} illustrates the result of our calculations in the 
  Hertzsprung-Russell diagram, encompassing remnant masses from well below
  0.2 up to above 0.4~M$_{\odot}$. The two sequences between 0.2 and 
  0.3~M$_{\odot}$ experienced typical thermal instabilities of thin burning shells
  when the CNO cycle shuts off (cf.\ Kippenhahn et al.\ 1968, Iben \& Tutukov 
  1986, Castellani et al.\ 1994). The latter authors find CNO flashes for
  masses above 0.3 M$_{\odot}$ only for Pop.\ II compositions. 
  
\section{Structures and cooling properties}

  We found a steep correlation between the remnant masses and the sizes of
  their hydrogen-rich envelopes, ranging from $5\cdot 10^{-2}$ M$_{\odot}$ for our
   0.179 M$_{\odot}$ model, down to $2\cdot 10^{-3}$ M$_{\odot}$ for
   0.414 M$_{\odot}$. Our envelope masses agree, for the mass range
  in common, $ M > 0.3$ M$_{\odot}$, with those given in Castellani et al.\ 
  (1994). For $ M < 0.2$ M$_{\odot}$  they agree in mass 
  {\em and\/} helium enrichment with those of the Sarna et al.\ (1998) models.
  It should be emphasized that these evolutionary envelope masses are  
  larger, for a given
  remnant mass, than those adopted recently by Benvenuto \& Althaus (1998) and
  Hansen \& Phinney (1998). 
  
\begin{figure}[ht]                                     
\plotone{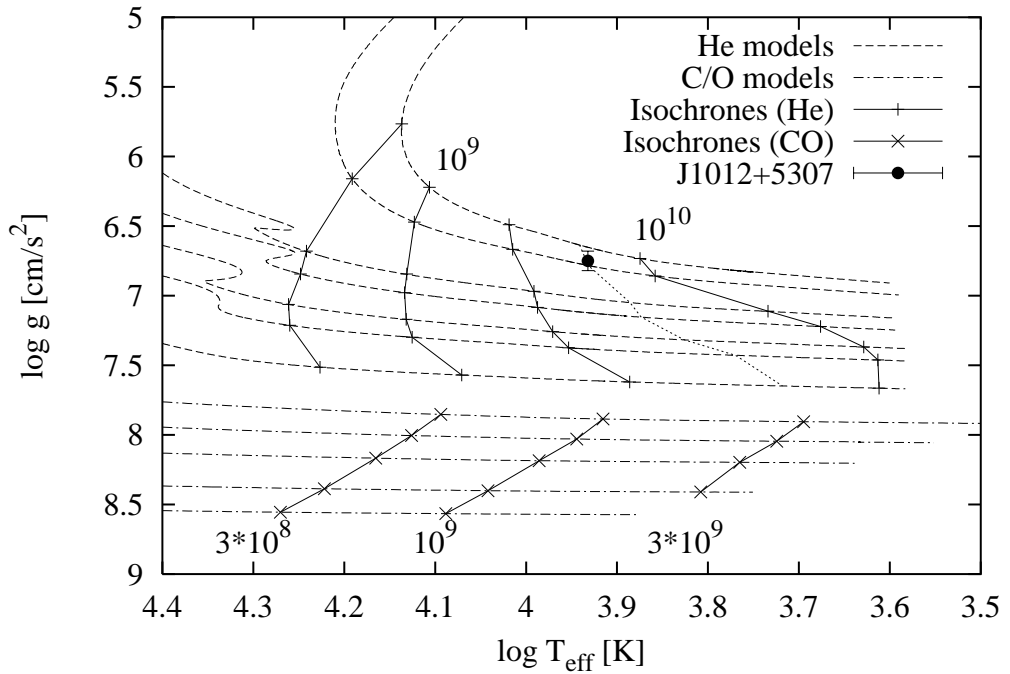}
  \caption{ \label{fig4}    
           $\log g-\log T_{\rm eff}$ diagram with evolutionary tracks of white 
           dwarfs with different masses (from top) of 0.179, 0.195, 0.234, 0.259,
           0.300, 0.331, 0.414 M$_{\odot}$ with helium cores, and of  0.524, 
           0.605, 0.696, 0.836, 0.940 M$_{\odot}$ with carbon-oxygen cores. 
           Isochrones are given between 0.3 and 10 Gyr, and the position of the
           PSR J1012+5307 companion is also indicated (van Kerkwijk et 
           al.\ 1996), together with the 6~Gyr isochrone (dotted).
          }
\end{figure} 
  Hydrogen burning continues to increase the helium core at the expense of
  the envelope, and the pp cycle 
  takes over on the cooling branch and completely determines the cooling rate 
  for the lower-mass models (Webbink 1975, Castellani et al.\ 1994). The
  continued burning at the base of the envelope reduces its mass considerably
  along the cooling path, leading to a complicated dependence of the white
  dwarf's size on total mass and effective temperature. Using our evolutionary
  mass-radius relationships instead of the ones available in the literature, 
  larger white dwarf masses would follow for given  surface gravities and 
  effective temperatures (see Driebe et al.\ 1998 for more details).   

  At the hot end of the
  cooling branch the hydrogen luminosity can exceed the gravothermal 
  contribution to the white dwarf's energy  budget by up to a factor of 100. 
  Even at effective temperatures as low as  5\,000~K the pp cycle still
  dominates in the models with $ M < 0.2$~M$_{\odot}$. 
  Since the evolution along the cooling sequence is slowed down accordingly,
  the isochrones for helium dwarfs differ in shape from those for CO dwarfs:
  they are shifted and turned over to the left (Fig.~\ref{fig4}).   

  The position of the  PSR J1012+5307 white-dwarf companion is met by our
  0.195 M$_{\odot}$ track for a cooling age of 6 Gyr, which
  agrees well with the spin-down age estimate of the pulsar, 7 Gyr. It should be 
  mentioned that this result is in close agreement with that of Sarna et al.\ 
  (1998) who made a  rather detailed modeling of this particular system (see
  also Alberts et al.\ 1996).

\section{Comparison with non-evolutionary models}

\begin{figure}[ht]                                      
\plotone{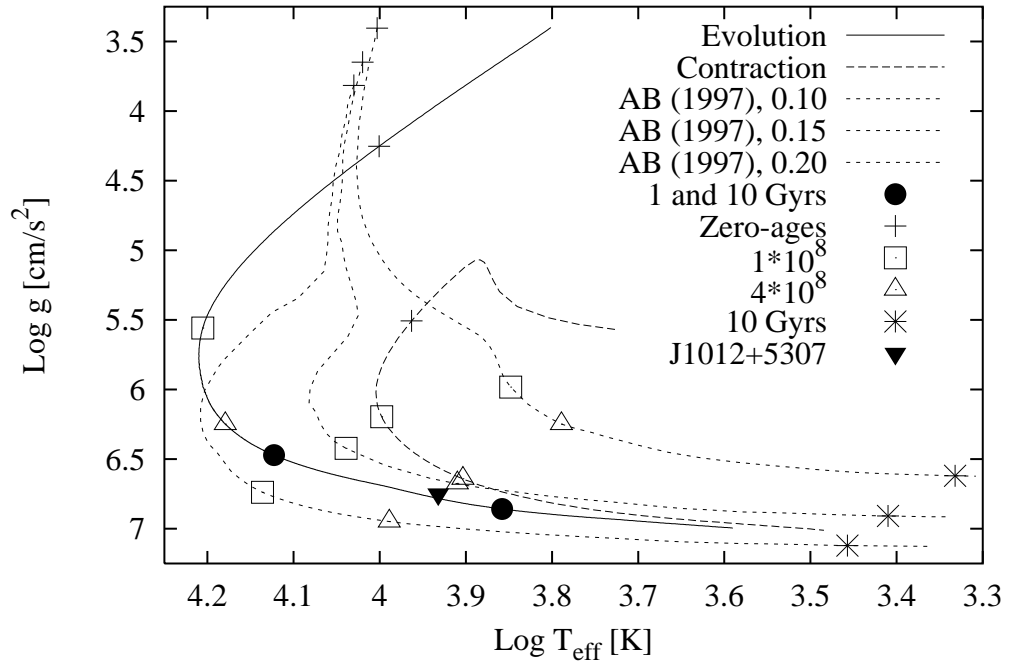}
  \caption{\label{fig5} 
           $\log g-\log T_{\rm eff}$ diagram with a 0.195~M$_{\odot}$ 
           evolutionary and non-evolutionary white-dwarf track as computed by
           us with identical chemical structures, and with three 
           non-evolutionary tracks taken from Althaus \& Benvenuto (1997). 
           Selected evolutionary ages are marked, and the
           position of the PSR J1012+5307 companion as well.
          }     
\end{figure}
  It is very instructive to compare the behaviour of our evolutionary models
   with that of non-evolutionary ones, as is done in Fig.~\ref{fig5} 
   (see also Bl\"ocker et al.\ 1997). 
   Despite the completely different initial structures of the two sets of
   non-evolutionary models shown in the figure, and somewhat 
   different physical assumptions as well, the cooling properties are
   identical, and the models predict an age of only 0.4 Gyr for the 
   PSR J1012+5307  companion, in variance with the pulsar's spin-down age. Note
   also that the structure of our non-evolutionary model does not approach that of the
   evolutionary model before an effective temperature of about 5\,000 K is
   reached.

  From a thorough comparison between evolutionary and non-evolutionary 
  models (cf. Bl\"ocker et al.\ 1997) we can make
  the following safe conclusions:
\begin{itemize}       \item
      Evolutionary white dwarf models are more compact than non-evolutionary
      models of the same mass and chemical structure.
                      \item
      Envelope masses are inversely correlated with the  white-dwarf mass.
                      \item
      At lower masses, hydrogen burning via the pp cycle controls the pace
      of cooling.
                      \item
      The thermo-mechanical structures of low-mass non-evolutionary models do 
      not  converge  with those of evolutionary models within a 
      reasonable time.  
\end{itemize} 
Given these facts, the use of non-evolutionary helium white-dwarf models to 
interpret observational data appears  not to be advisable.

\acknowledgments

F.H. and T.B. thank the DFG for financial support (grants Scho 394/13 and 
                                                            Ko 738/12).

\end{document}